\documentclass[aps,prl]{revtex4}
\usepackage{graphicx}
\begin{document}
\title{Solvent content of protein crystals from diffraction intensities 
by Independent Component Analysis}

\author{Massimo Ladisa$^a$}
\affiliation{Istituto di Cristallografia--CNR, c/o Dipartimento Geomineralogico, 
Universit\`a degli Studi di Bari, via E. Orabona 4, I-70125 Bari, Italy}
\author{Antonio Lamura}
\affiliation{Istituto Applicazioni Calcolo--CNR, sezione di Bari, 
via G. Amendola 122/D, I-70126 Bari, Italy}
\author{Giovanni Nico}
\affiliation{Istituto di Radioastronomia--CNR, sezione di Matera, 
c/o Centro di Geodesia Spaziale--ASI, I-75100 Matera, Italy}
\author{Dritan Siliqi}
\affiliation{Faculty of Natural Sciences, Department of Inorganic Chemistry, 
Laboratory of X--ray Diffraction, Tirana 5, Albania}
\date{\today}
\vskip 0.5cm
\begin{abstract}
An analysis of the protein content of several crystal 
forms of proteins has been performed. We apply a new numerical technique, the 
Independent Component Analysis (ICA), to determine the volume fraction of 
the asymmetric unit occupied by the protein. 
This technique requires only the crystallographic data of structure factors
as input. 
\end{abstract}
\maketitle
\noindent
$^a$ Corresponding author,
E\_mail: massimo.ladisa@ic.cnr.it,
Phone: +39 0805442419,
Fax: +39 0805442591\\
\noindent
SHORT TITLE: ICA in protein crystallography

\section{Introduction}

Protein crystals contain between around 30\% and 70\% of solvent \cite{matthewsJMB1968}, 
most of which is disordered in the solvent channels among the protein molecules 
of the crystal lattice. Thus the electron densities of the protein molecules, 
with typical values of 0.43 e/\AA$^3$, are surrounded by a continuous disordered 
solvent electron density ranging between 0.33 e/\AA$^3$ for pure water and 
0.41 e/\AA$^3$ for 4M ammonium sulphate \cite{kostrewaCCP42002}.

If we do not account for any  model for this continuous disordered solvent 
electron density, atomic protein models are thought as if it were placed in vacuum. 
The electron density itself is overestimated, the calculated structure factor 
amplitudes are systematically much larger than the observed ones \cite{kostrewaB1995}, 
and it is commonly believed that the latter condition especially occurs at low 
resolution. 

The higher is the discrepancy among the calculated structure factors and the 
observed ones, the more difficult is the data scaling, the least-square refinement 
and the electron density map rendering. Cutting low resolution data has been a 
widely adopted method to step over the problem, although it was rough and, somehow, 
intrinsically wrong: Indeed, it introduced distortions of the local electron density 
(an optical example is discussed in \cite{cowtan}). 

A great effort has been recently devoted to devise a reliable method accounting for 
the disordered solvent effects in the protein region \cite{anderssonACD2000,liACD2003}. 
Two of them deserve a brief 
description.  
\begin{itemize}
\item[I.]
The exponential scaling model \cite{moewsJMB1975}.
\par
\noindent
This model is obtained by the direct application of Babinet's principle to the 
calculated structure factors. The solvent structure factors moduli are assumed 
to be proportional to the protein ones, whereas the phases are opposite. The 
lower is the resolution the more satisfactory is the agreement between the 
observed structure factor and the calculated one \cite{urzCCP41995}. Due to 
its simplicity, this model has been implemented in most of the crystallographic 
refinement programs \cite{dodsonCCP41996}. The weakness of this model is strictly 
related to the resolution at which it is expected to work properly. Indeed the 
approximation embodied in this method is true at resolutions below $\approx$ 15\AA\ 
although it can be stretched up to $\approx$ 5\AA\ by downscaling the structure factors. 
\item[II.]
The mask model \cite{jiangJMB1994}.
\par
\noindent
This model is an improvement of the previous one, since it aims to sum up the 
protein structure factor and the solvent one vectorially, {\it i.e.} by 
accounting for both the modulus and the phase of the two structure factors. 
In the mask model the protein molecules are placed on a grid in the unit cell 
whereas the grid points outside the protein region are filled with the disordered 
solvent electron density. The protein boundary is mainly determined by the 
Van--der--Waals radii. The disordered solvent electron density is 
$^{\prime\prime}$stretched$^{\prime\prime}$ to fill in the empty space and the 
calculation of the solvent structure factor is straightforward. Although the 
mask model works rather well, there are three major drawbacks of it: Too many 
parameters have to be fitted and the relatively large $parameters/observables$ 
ratio weakens the model at high resolution (overfitting); finally, 
the disordered solvent 
electron density is unrealistically assumed to be step shaped and flat. Some 
strategies have been already devised to improve the latter ones \cite{jiangJMB1994}.  
\end{itemize}
\vskip20pt
The aim of this paper is to focus on a recently developed statistical method 
and its application to disentangle the protein and the solvent contributions out 
of crystallographic data; we show its major advantages and drawbacks. Up to our knowledge 
this method has never been applied to this field. A comparison of the protein fraction 
in the unit cell, calculated by this method, with the same quantity computed 
by the most popular method used nowadays \cite{matthewsJMB1968} is satisfactory. 
\par
\noindent
The plan of the paper is as follows: The {\it theory} section provides the reader with 
the basic concepts of the independent component analysis; the 
{\it results and discussion} section applies the theory to the specific case 
of a 2-dimensional problem ({\it i.e.} solvent/protein system) we are interested in; 
it concludes with the calculation of the protein fraction for several protein structures 
and with a comparison of this quantity with the analogous one calculated by 
the Matthews' model accounting for the protein content only. 
{\it Conclusions} section summarizes the paper's content and suggests further 
investigations.

\section{Theory}

Several techniques have been devised so far to deal with protein 
crystallography. Among them we quote the isomorphous derivative 
(SIR, MIR) and the anomalous dispersion (SAD, MAD) ones 
(we address the reader to a number of review papers for details on 
these techniques; see, for instance, \cite{giacovazzoDPC1998} and 
references therein).
\par
\noindent
The theory described hereafter can be applied to protein crystallography 
regardless of the specific technique we are using and without any substantial 
modification; therefore, for the sake of simplicity, we shall focus on the 
isomorphous derivative one. Anyway the method will be finally applied to 
several proteins: Among them some are anomalous dispersion structures 
and some others refer to the isomorphous derivative technique.
\vskip10pt
\par
A protein and its isomorphous derivatives crystallize in a solvent. 
Imagine that you are measuring the diffraction intensities out of a 
crystallized protein sample and one of its isomorphous derivatives. 
Each of these recorded signals is a weighted sum of the signals 
emitted by the two main sources 
({\it i.e.} protein/derivative and solvent), which we denote by 
${\mathcal F}^{p/d}$ and ${\mathcal F}^s$, {\it i.e.} the 
protein/derivative and solvent structure factors, respectively. 
We can express each of them 
as a linear combination
\begin{eqnarray}
{\mathcal F}^{p+s} &=& a^{p+s}_p~ {\mathcal F}^p + a^{p+s}_s~ {\mathcal F}^s \;\; , \nonumber \\
{\mathcal F}^{d+s} &=& a^{d+s}_d~ {\mathcal F}^d + a^{d+s}_s~ {\mathcal F}^s \;\; .
\label{lin-comb-F}
\end{eqnarray}
Actually if we knew the $a^i_j$ parameters we would solve the problem at 
a once by classical methods. Unfortunately this is not the case and the 
problem turns out to be much more difficult. 
\par
\noindent
Under the hypothesis of statistical independence of the ${\cal F}^{p,s}$ structure factor  
phase differences, {\it i.e.} $\langle \cos(\phi^s - \phi^p)\rangle \simeq 0$ \cite{kostrewaCCP42002}, 
we can write
\begin{eqnarray}
{\mathcal I}^{p+s} &=& a^{p+s\ 2}_p~ {\mathcal I}^p + a^{p+s\ 2}_s~ {\mathcal I}^s \;\; , \nonumber \\
{\mathcal I}^{d+s} &=& a^{d+s\ 2}_d~ {\mathcal I}^d + a^{d+s\ 2}_s~ {\mathcal I}^s 
~\approx~ a^{d+s\ 2}_d~ {\mathcal I}^p + a^{d+s\ 2}_s~ {\mathcal I}^s \;\; ,
\label{lin-comb}
\end{eqnarray}
where ${\mathcal I}^{p,s} \propto \langle |{\mathcal F}^{p,s}|^2 \rangle$ is the resolution shell 
averaged intensity and  
the approximation in the last equation written above is justified 
by the isomorphism among the protein and its derivatives. 
$a^i_j$ are some parameters that depend on the hidden variables of 
the problem. Of course we are interested in spotting the two original 
sources ${\mathcal I}^p$ and ${\mathcal I}^s$ by using only the 
recorded signals ${\mathcal I}^{p+s}$ and ${\mathcal I}^{d+s}$.

Using some information about the statistical properties of the original 
signals ${\mathcal I}^p$ and ${\mathcal I}^s$ is a possible approach to 
estimate the $a^i_j$ parameters. The {\it statistical independence} 
of the two sources is not surprising whereas the fact that the above 
condition is not only necessary but also sufficient is 
\cite{hyvarinenNN2000}. 

The Independent Component Analysis is a technique recently 
developed to estimate the $a^i_j$ parameters based on the information 
of the statistical independence of the original sources. It allows to 
separate the latter ones from their mixtures ${\mathcal I}^{p+s}$ and 
${\mathcal I}^{d+s}$.
 
Several applications of ICA have been recently devised and, therefore, 
a unified mathematical framework is required. 

To begin with, we rigorously define ICA \cite{comonSP1994,juttenSP1991} by 
referring to a statistical $^{\prime\prime}$latent variables$^{\prime\prime}$ 
model, {\it i.e.}
\begin{equation}
x_j \stackrel{{\it def.}}{=} a_j^1~s_1 + a_j^2~s_2 +~...~+ a_j^n~s_n  \;\; ,
\label{ICA-def}
\end{equation}
where j runs over the number of linear mixtures we observe and n is the number 
of hidden sources. 
\par
\noindent
The statistical model defined in eq.(\ref{ICA-def}) is called Independent 
Component Analysis. It describes the generation of observed 
data $x_j$ as a result of an unknown mixture $a_j^i$ of unknown sources 
$s_i$. Finding out both the mixing matrix and the hidden sources is the aim 
of this method. In order to do so, ICA assumes that
\begin{itemize}
\item[1)]
the components $s_i$ are statistically independent,
\item[2)]
the components $s_i$ are random variables and their distribution is 
\underline{not} gaussian,
\item[3)]
the mixing matrix $a_j^i$ is square, although this hypothesis can be sometimes relaxed. 
For a detailed discussion see \cite{hyvarinenNN2000}.
\end{itemize}
Let us suppose that the mixing matrix $a_j^i$ has been computed; the inverse mixing 
matrix $w^j_i$ is achievable and the problem is readily solved: 
$\displaystyle{s_i}~=~w^1_i~x_1 + w^2_i~x_2+~...~+ w^n_i~x_n$ for each hidden source.
\par
\noindent
Adding some noise terms in the measurements is certainly a more realistic approach 
although it turns out to be more tricky: For the time being, we shall skip this aspect 
in order to focus on a free-noise ICA model. Of course extending the conclusions to more 
complicated models is straightforward.
\par
\noindent
Without loss of generality we shall assume that $x_j$ are standardized random variables, 
{\it i.e.} $\displaystyle{Variance(x_j)=1}$, $\displaystyle{Mean(x_j)=0}$. The latter 
choice is always possible since both $Variance$ and $Mean$ are known for the starting 
data samples $x_j$. Indeed, we can always replace the starting set of random variables 
$x_j$ with the new one as follows  
\begin{equation}
\tilde x_j \stackrel{{\it def.}}{=} \frac{x_j - Mean(x_j)}{\sqrt{Variance(x_j)}} \;\; .
\label{STD-vars}
\end{equation}
Moreover ICA aims to disentangle the hidden sources ($s_i$) and, therefore, looking 
at preprocessing techniques to uncorrelate the {\it would-be} sources before 
applying any ICA algorithm is a major advantage. Therefore this procedure, named 
data whitening, is certainly a useful preprocessing strategy in ICA. The eigenvalue 
decomposition (EVD) is the most popular way to whiten data: The starting set of 
variables is linearly transformed according to the following rule 
\begin{equation}
\dot x_j \stackrel{{\it def.}}{=} 
\left(\frac{V_j^1~V_1^1}{\sqrt{\lambda_1}} + 
\frac{V_j^2~V_2^1}{\sqrt{\lambda_2}} +~...~+ 
\frac{V_j^n~V_n^1}{\sqrt{\lambda_n}}\right)~x_1 +~...~+ 
\left(\frac{V_j^1~V_1^n}{\sqrt{\lambda_1}} + 
\frac{V_j^2~V_2^n}{\sqrt{\lambda_2}} +~...~+ 
\frac{V_j^n~V_n^n}{\sqrt{\lambda_n}}\right)~x_n  \;\; ,
\label{whitened-vars}
\end{equation}
where $\displaystyle{\lambda_j}$ and $\displaystyle{\{V_j^i\}_{i=1,...,n}}$ are, 
respectively, the eigenvalues and the eigenvectors of the n-rank covariance matrix 
for the starting set of statistical variables. The covariance matrix for the new 
set of $\dot x_j$ variables is diagonal. 
\par
\noindent
Of course the data whitening modifies the mixing matrix $a_j^i$; infact, by 
applying the ICA definition of the eq.(\ref{ICA-def}) to both sets of variables 
($x_j$ and $\dot x_j$) in the eq.(\ref{whitened-vars}), we get
\begin{equation}
\dot a_j^i \stackrel{{\it def.}}{=} 
\left(\frac{V_j^1~V_1^1}{\sqrt{\lambda_1}} + 
\frac{V_j^2~V_2^1}{\sqrt{\lambda_2}} +~...~+ 
\frac{V_j^n~V_n^1}{\sqrt{\lambda_n}}\right)~a_1^i +~...~+ 
\left(\frac{V_j^1~V_1^n}{\sqrt{\lambda_1}} + 
\frac{V_j^2~V_2^n}{\sqrt{\lambda_2}} +~...~+ 
\frac{V_j^n~V_n^n}{\sqrt{\lambda_n}}\right)~a_n^i  \;\; ,
\label{whitened-mix-mat}
\end{equation}
where i runs over 1,...,n. It turns out that data whitening has considerably 
simplified the initial problem since the new mixing matrix $\dot a_j^i$ is 
orthogonal and, therefore, the n$^2$ components of $a_j^i$ have been reduced 
to $\displaystyle{n(n-1)/2}$.
\vskip10pt
\par
After having standardized and whitened the starting set of statistical 
variables, we are ready to implement the ICA algorithm. 
\par
\noindent
Actually we are looking for a unique matrix $\dot w_i^j$ that combines with 
the $\dot x_j$ variables in order to get the hidden sources $\dot s_i$ satisfying 
the ICA prescriptions. Indeed the conditions $1),3)$ of ICA are readily achieved 
as soon as we note that the product of the whitened set of standardized random 
variables $\dot x_j$ by any n-dimensional orthogonal matrix leaves the variables 
uncorrelated, whitened, standardized and, moreover, it leaves the mixing matrix 
$\dot a_j^i$ orthogonal. Therefore we shall limit ourselves to an n-dimensional 
orthogonal matrix $\dot w_i^j$ and we shall fix its $\displaystyle{n(n-1)/2}$ 
degrees of freedom by assuming the nongaussianity of the probability distribution 
functions of the hidden sources 
$\displaystyle{\dot s_i}~=~\dot w^1_i~\dot x_1 + \dot w^2_i~\dot x_2+~...~+ 
\dot w^n_i~\dot x_n$.
\vskip10pt
\par
There are several measures of nongaussianity and a full discussion is 
beyond the scope of this paper (for more details see \cite{hyvarinenNN2000}); 
instead we briefly introduce the measure of nongaussianity we shall adopt: 
The negentropy. 
\par
\noindent
Entropy is a fundamental concept of information theory. The entropy of a random 
variable is its coding length (for details see \cite{cover1991,papoulis1991}). 
For a discrete random variable, entropy is defined as follows
\begin{equation}
\mathcal{H}(Y) \stackrel{{\it def.}}{=}-\sum_i~P(Y=\xi_i)~log~P(Y=\xi_i) \;\; ,
\label{entropy}
\end{equation}
where $\xi_i$ are the possible values of $Y$. One of the main result of the 
information theory is that a gaussian variable has the largest entropy 
among the random variables with the same $Variance$. Therefore we argue that 
the less structured is a random variable the more gaussian is its distribution. 
In order to get a nonnegative measure of a random variable nongaussianity, whose 
value is zero for a gaussian variable, it is worth to introduce the following 
quantity
\begin{equation}
{\mathcal{J}}(Y) \stackrel{{\it def.}}{=} \mathcal{H}(Y_{gauss.})-\mathcal{H}(Y) \;\; ,
\label{negentropy}
\end{equation}
where $\mathcal{H}(Y_{gauss.})$ is the entropy of a gaussian random variable. 
Hereafter we shall refer to the eq.(\ref{negentropy}) as to the negentropy of a 
random variable $Y$. 
\vskip10pt
\par
The negentropy of a random variable, as defined in the 
eq.(\ref{negentropy}), is well defined by the statistical theory and, moreover, 
it can be easily generalized to a system of random variables: Infact the additivity 
of the entropy is immediately extended to the negentropy. Moreover negentropy is 
invariant under an invertible linear transformation \cite{comonSP1994,hyvarinenNCS1999}. 
The major drawback of the negentropy, as defined in the eq.(\ref{negentropy}), is 
the computation itself since the precise evaluation of it requires the 
nonparametric estimation of the probability distribution function for the random 
variable we are dealing with. Several simplifications of negentropy have been 
devised and we shall focus on two of them. 
\begin{itemize}
\item[i.] The kurtosis \cite{jonesJRSS1987}.
\par
\noindent
The kurtosis is the 4$^{th}$ order momentum of a random variable probability 
distribution function, {\it i.e.} 
$\displaystyle{kurtosis(Y)\stackrel{{\it def.}}{=}Mean(Y^4)/Mean(Y^2)^2}-3$, 
where $Y$ is a random variable. For a gaussian random variable kurtosis equals 
0. The negentropy of eq.(\ref{negentropy}) is readily simplified: 
$\displaystyle{\mathcal{J}(Y)\approx \frac{1}{12} Mean(Y^3)^2+\frac{1}{48} kurtosis(Y)^2}$. 
\par
\noindent
The major drawback of the kurtosis approximation of negentropy is the lack of robustness, 
since its computation out of a data sample can be very sensitive to the outliers 
\cite{huberTAS1985}.
\item[ii.] Maximum entropy \cite{hyvarinenANIPS1998}.
\par
\noindent
In order to step over the unrobustness of the negentropy approximation described above, 
it is useful to introduce a conceptually simple and fast to be computed approximation 
of the negentropy based on maximum entropy principle. We write the negentropy 
according to the following formula
\begin{equation}
\label{negentropy-approx}
\mathcal{J}(Y)\approx 
\sum_{j=1}^N~c_j\left[Mean(~C_j(Y)~)-Mean(~C_j(Y_{gauss.})~)\right]^2 \;\; ,
\end{equation} 
where $c_j$ are suitable coefficients, $C_j$ are nonquadratic functions, $Y$ is a unit 
variance random variable and $Y_{gauss.}$ is a unit variance gaussian random variable. 
The approximation of eq.(\ref{negentropy-approx}) generalizes the kurtosis one; infact 
for a single function $C_j$ ({\it i.e.} N=1) the choice $C_1=Y^4$ exactly leads to the 
kurtosis approximation described above. The slower is the growing of the $C_j$ functions 
the more robust is the approximation of the negentropy.  
\end{itemize}
Both of the approximations described above satisfy the main features of the negentropy, 
{\it i.e.} the nonnegativity, the zero value for a gaussian random variable and the 
additivity. 
\vskip10pt
\par
Before showing the details of the ICA application to the protein crystallography, 
we spot some intrinsic ambiguities of the ICA procedure \cite{hyvarinenNCS1999}. 
\begin{itemize}
\item[a)] The $Variance$ of the independent components cannot be determined since 
the hidden sources and the mixing matrix are unknown and they can be 
simultanously scaled by the same quantity without modifying any conclusion. 
The choice $Variance(Y)=1$ leaves the ambiguity of the sign.
\item[b)] The order of the independent components cannot be determined. Indeed any 
permutation of the hidden sources leads to a similarity transformation on the 
mixing matrix and since both of them are unknown the permutation does not affect 
the algorithm itself. 
\end{itemize}
The ambiguities described above can be solved by means of physical contraints 
featuring the ICA solutions of the specific problem. We will discuss how to overcome 
this problem in the next section. 

\section{Results and discussion}
We have focused on the 2-dimensional problem described in the introduction and 
briefly formalized at the beginning of the theory section.
\par
\noindent 
After having recalled the definitions of the eq.(\ref{lin-comb}), we proceed with 
the standardizing and the withening procedures of the starting set of random 
variables
\begin{equation}
I^{p/d+s} 
\ \ \ \stackrel{standardizing}{\longrightarrow} \ \ \ 
\tilde I^{p/d+s} 
\ \ \ \stackrel{whitening}{\longrightarrow} \ \ \ 
\dot I^{p/d+s} \stackrel{{\it def.}}{=} 
\left[ \mathcal{A}_{wh.} \cdot \tilde I \right]^{p/d+s} \;\; ,
\label{STD-whitening}
\end{equation}
where $\mathcal{A}_{wh.}$ is the whitening matrix defined by the eigenvalue decomposition 
(EVD) of the $\tilde I$ $Covariance$ matrix. 
\vskip10pt
At this stage we apply the ICA algorithm to $\dot I^{p/d+s}$. In two dimensions an 
orthogonal matrix is determined by a single angle parameter; we get
\begin{equation}
\mathcal{A}_{ICA}(\theta)\stackrel{{\it def.}}{=}
\left(
\begin{tabular}{c c}
$\cos \theta$ & $-\sin \theta$ \\
$\sin \theta$ & $\cos \theta$
\end{tabular}
\right) \longrightarrow 
\left(
\begin{tabular}{c} 
$\dot I^p(\theta)$ \\ 
$\dot I^s(\theta)$ 
\end{tabular}
\right)
\stackrel{{\it def.}}{=} 
\mathcal{A}_{ICA}(\theta) \cdot \mathcal{A}_{wh.} \cdot
\left(
\begin{tabular}{c}  
$\tilde I^{p+s}$ \\ 
$\tilde I^{d+s}$
\end{tabular}
\right) \;\; ,
\label{ICA-matrix}
\end{equation}
where the last formula of the eq.(\ref{ICA-matrix}) defines the 
$\theta$-dependent solutions of ICA, {\it i.e.} the standardized, 
whitened random variables depending on the single parameter $\theta$ 
that has to be fixed by maximizing the total negentropy $\mathcal{J}(\theta)$ 
as follows
\begin{equation}
\mathcal{J}(\theta)\stackrel{{\it def.}}{=}
\mathcal{J}(\dot I^p(\theta))+\mathcal{J}(\dot I^s(\theta)) 
\ \ \ \ maximum \;\; ,
\label{max-negentropy}
\end{equation} 
where we use a single function $C_j$ ({\it i.e.} N=1 in the 
eq.(\ref{negentropy-approx})) and, according to 
\cite{hyvarinenNN2000}, we adopt $\displaystyle{C_1(Y) = -\exp(-Y^2/2)}$. 
In the eq.(\ref{max-negentropy}) the negentropy additivity has been applied. 
Other choices for $C_1$ are possible \cite{hyvarinenNN2000} and we have 
checked that neither the solutions nor the algorithm are sensitive to them. 
\par
\noindent
We denote with $\theta_{max}$ the angle $\theta$ where the total negentropy 
$\mathcal{J}(\theta)$ attains its maximum. Therefore we can conclude 
\begin{equation}
\mathcal{A}_{ICA}(\theta_{max}) = 
\left(
\begin{tabular}{c c}
$\displaystyle{\dot a^{p+s}_p}$ & $\displaystyle{\dot a^{p+s}_s}$ \\
$\displaystyle{\dot a^{d+s}_p}$ & $\displaystyle{\dot a^{d+s}_s}$
\end{tabular}
\right)^{-1} \;\; ;
\label{A^-1}
\end{equation}
in that respect $\dot I^{p/s}(\theta_{max})$ are the standardized, whitened and 
maximally nongaussian random variables corresponding to the hidden sources 
of the initial problem. 
\par
As to the ambiguities of this technique, mentioned at the end of the previous 
section, we solve the first one by taking the absolute value of 
$\dot I^{p/s}(\theta_{max})$, {\it i.e.} we introduce the quantities 
${\displaystyle \mathcal{I}^{p/s}}\stackrel{{\it def.}}{=}|\dot I^{p/s}(\theta_{max})|$.
\par
At this stage we define the protein/solvent fraction as follows
\begin{equation}
f^{p/s} \stackrel{{\it def.}}{=} 
\frac{ \sum_j~ \Delta\rho_j~ \mathcal{I}^{p/s}_j }
{ \sum_j~ \Delta\rho_j~ \left[~ \mathcal{I}^{p}_j+\mathcal{I}^{s}_j ~\right] } \;\; ,
\label{prot-frac}
\end{equation}
where $j$ runs over the number of resolution shells according to eq. (\ref{lin-comb}) and 
$\displaystyle{ \Delta\rho_j~=~ \langle \rho \rangle_{j+1} - \langle \rho \rangle_{j} }$, 
being $\langle \rho \rangle$ the shell averaged resolution.
\par
The protein fraction 
definition of eq.(\ref{prot-frac}) is justified by the kinematic 
theory \cite{authierDTXD2001} stating that the 
diffraction intensity is expected to depend on the crystal volume 
$\Omega$ and on the unit cell volume V according to the 
$\displaystyle \Omega/V^2$ ratio. 
\par
According to \cite{bricogneACD2003} 
the relevant information of the protein structures is contained in three resolution ranges, 
$\leq 1.2$\AA, $1.7-3.0$\AA~ and $\geq 3.5$\AA. The first range information is dominated by 
the protein structure at atomic level while in the third one the solvent content is overwhelming
(the Density Modification procedures aim to re-scale the structure factors moduli in the 
low resolution range to account for the bulk disordered solvent). The 
scattering powers of protein and solvent are of the same order in the $1.7-3.0$\AA~ range 
\cite{kostrewaCCP42002}. Hence the expression (\ref{prot-frac}) is evaluated in this range.
\par
\noindent
The comparison between the protein fraction value obtained by ICA with 
the one by Matthews' method \cite{matthewsJMB1968}, computed as in 
\cite{kantardjieffPS2003}, 
finds out the correct 
order of the independent components ${\displaystyle \mathcal{I}^{p/s}}$. 
\vskip10pt
The results of this comparison are shown in Table \ref{tab:fp} for several proteins.
The agreement between the protein fractions obtained by the two methods is quite satisfactory. 
\par
\noindent
The proteins reported in Table \ref{tab:fp} are named according to 
their codes. 
For each of them we have the crystallographic data of the native and of 
one derivative for the isomorphous derivative technique. 
For the anomalous dispersion 
technique, we use the crystallographic data of the native collected at 
one wavelength. 
\par
On the last row in the Table \ref{tab:fp} the errors are 
the protein fraction $Variances$ for the two different methods.
The average values as well as the $Variances$ are comparable.
\par
\begin{figure}[ht!]
\begin{center}
\includegraphics[width=0.45\textwidth]{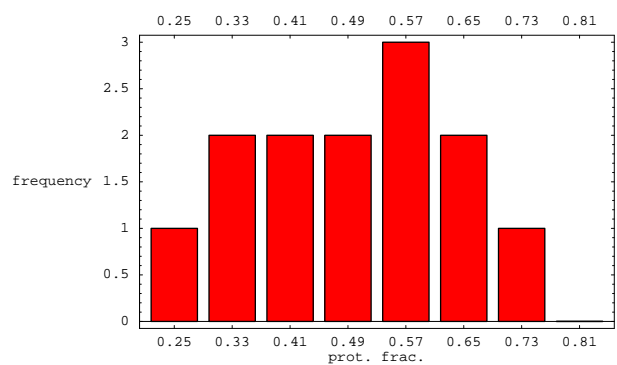}
\caption{Protein fraction distribution for the proteic 
structures listed in Table \ref{tab:fp} computed by ICA.}
\label{fig:crys-vol1}
\end{center}
\end{figure}
\begin{figure}[ht!]
\begin{center}
\includegraphics[width=0.45\textwidth]{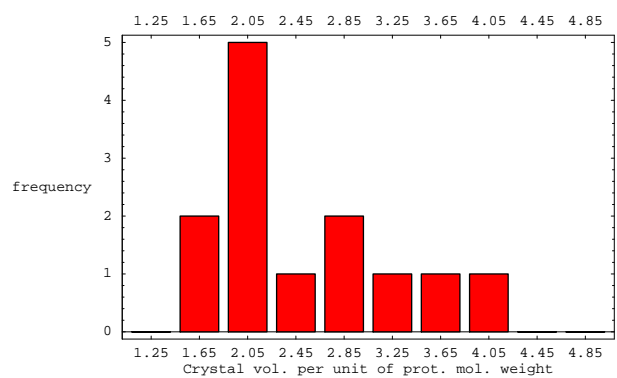}
\caption{Crystal volume per unit of protein molecular weight 
distribution for the proteic structures listed in 
Table \ref{tab:fp}
computed by ICA (for a comparison see Fig.2 in ref.\cite{matthewsJMB1968}). 
x-axis units are \AA$^3$/Dalton.}
\label{fig:crys-vol2}
\end{center}
\end{figure}

In Fig.\ref{fig:crys-vol1} we report the protein fraction distribution 
for the proteic structures listed in Table \ref{tab:fp} and computed by ICA. 
Figure \ref{fig:crys-vol2} shows the corresponding distribution of the crystal volume 
per unit of protein molecular weight calculated 
according to the formula in ref.\cite{matthewsJMB1968}
(for a full comparison see Fig.2 in the reference \cite{matthewsJMB1968}).  
According to our analysis the most probable value for the crystal volume 
per unit of protein molecular weight falls into the range 1.85-2.25 \AA$^3$/Dalton.
\vskip10pt
\section{Conclusions}
In this paper we have applied a new technique, the Independent Component Analysis,
to calculate the protein fraction 
out of crystallographic data. The analysis here presented
aims to disentangle the protein and the disordered solvent 
contributions. Provided a sufficient number of crystallographic data (at least as 
many as the supposed hidden sources), this method has given convincing results, as
compared to available ones in the literature. It is a promising 
tool to investigate some features of protein structures even if its applicability 
as a robust guideline at the future protein crystallography refinement programs 
deserves a deeper investigation.
\vskip10pt
\par
We quote some possible directions of research:
\begin{itemize}
\item[1.] Phasing procedures. Indeed weighting the protein structure factors 
according to the resolution shells of the crystallographic data could provide 
a crucial improvement of the relevant formulas for the protein phasing procedure 
implemented in the most popular crystallography refinement programs. 
\item[2.] Disentangling crystallized and disordered solvent contributions out 
of the crystal forms of proteins. Infact the larger is the number of the 
independent crystallographic data referring to the same protein structure, 
the larger is the number of hidden sources this method can account for, the 
more precise is the determination of the single hidden source out of the 
recorded signals.
\item[3.] Model independence of ICA results in protein crystallography. 
\end{itemize}

\newpage
\begin{table}[ht!]
  \caption{{\small Numerical values for protein fractions. 
The third column refers to our method, the fourth column refers to Matthews' 
method \cite{matthewsJMB1968}. The protein fraction calculated by ICA is 
averaged on the whole crystallographic data resolution range. SIR, MIR, 
SAD and MAD refer to the diffraction technique adopted to collect data. 
On the last column we report the error estimate between the two methods.}}
\begin{center}
\begin{tabular}{|c|c|c|c|c|}
\hline 
$protein$ & $technique$ & $prot.~frac.$    & $prot.~frac.$ & $err.$ \\
          &    & $(~this~paper~)$ & $(~Matthews ~[1]~)$ & \\
\hline
$GMT~(Ortho)[24]$  & $SIR$             & $0.31$          & $0.30$          & $0.03$ \\
$GMT~(Mono)[24]$   & $^{\prime\prime}$ & $0.57$          & $0.53$          & $0.07$ \\
$SM_2[25]$         & $^{\prime\prime}$ & $0.68$          & $0.65$          & $0.05$ \\
$E_2[26]$         & $^{\prime\prime}$ & $0.33$          & $0.26$          & $0.24$ \\
$TXN[27]$        & $^{\prime\prime}$ & $0.44$          & $0.45$          & $0.02$ \\
$GLPE[28]$  & $^{\prime\prime}$ & $0.69$          & $0.61$          & $0.12$ \\
$APP[29]$          & $^{\prime\prime}$ & $0.65$          & $0.67$          & $0.03$ \\
$dUTPase[30]$   & $MIR$             & $0.40$          & $0.37$          & $0.08$ \\
$BPO[31]$           & $^{\prime\prime}$ & $0.45$          & $0.44$          & $0.02$ \\
$CAUFD[32]$        & $SAD$             & $0.79$          & $0.86$          & $0.08$ \\
$LYSO_2[33]$     & $^{\prime\prime}$ & $0.61$          & $0.58$          & $0.05$ \\
$KPR[34]$           & $MAD$             & $0.53$          & $0.54$          & $0.02$ \\
$NOX[35]$           & $^{\prime\prime}$ & $0.60$          & $0.51$          & $0.16$ \\
\hline
$average$                        &                   & $0.54 \pm 0.15$ & $0.52 \pm 0.16$ & $0.07$ \\
\hline
\end{tabular}
\end{center}
\label{tab:fp}
\end{table}
\newpage

\end{document}